\documentclass[conference]{IEEEtran}
\IEEEoverridecommandlockouts
\usepackage{cite}
\usepackage{amsmath,amssymb,amsfonts}
\usepackage{algorithmic}
\usepackage{graphicx}
\usepackage{textcomp}
\usepackage{xcolor}
\usepackage{float}
\ifCLASSOPTIONcompsoc \usepackage[caption=false,font=normalsize,labelfon
t=sf,textfont=sf]{subfig}
\else
\usepackage[caption=false,font=footnotesize]{subfi g}
\fi

\def\BibTeX{{\rm B\kern-.05em{\sc i\kern-.025em b}\kern-.08em
    T\kern-.1667em\lower.7ex\hbox{E}\kern-.125emX}}
    
\begin{document}

\title{Signal detection in extracellular neural ensemble recordings using higher criticism\\}

\author{\IEEEauthorblockN{Farzad Fathizadeh*}
\IEEEauthorblockA{\textit{Dept of Mathematics, Swansea University} \\
Swansea, United Kingdom \\
\textit{MPI for Biological Cybernetics}\\
T\"ubingen, Germany\\
farzad.fathizadeh@swansea.ac.uk}
\and
\IEEEauthorblockN{Ekaterina Mitricheva*}
\IEEEauthorblockA{\textit{MPI for Biological Cybernetics} \\
T\"ubingen, Germany\\
ekaterina.mitricheva@tuebingen.mpg.de}
\and
\IEEEauthorblockN{Rui Kimura}
\IEEEauthorblockA{\textit{MPI for Biological Cybernetics} \\
%\textit{}\\
T\"ubingen, Germany \\
rui.kimura@tuebingen.mpg.de}
\and
\IEEEauthorblockN{Nikos Logothetis}
\IEEEauthorblockA{\textit{MPI for Biological Cybernetics} \\
T\"ubingen, Germany \\
nikos.logothetis@tuebingen.mpg.de}
\and
\IEEEauthorblockN{Hamid R. Noori}
\IEEEauthorblockA{\textit{MPI for Biological Cybernetics} \\
T\"ubingen, Germany  \\
\textit{Courant Institute for Mathematical Sciences}\\
New York,  USA \\
hamid.noori@tuebingen.mpg.de}
\and
%\IEEEauthorblockN{ff}
%\IEEEauthorblockA{\textit{ff} \\
*These authors contributed equally. \\
}

\maketitle

\begin{abstract}
Information processing in the brain is conducted by a concerted action of multiple neural populations. Gaining insights in the organization and dynamics of such populations can best be studied with broadband intracranial recordings of so-called extracellular field potential, reflecting neuronal spiking as well as mesoscopic activities, such as waves, oscillations, intrinsic large deflections, and multiunit spiking activity.  Such signals are critical for our understanding of how neuronal ensembles encode sensory information and how such information is integrated in the large networks underlying cognition. The aforementioned principles are now well accepted, yet the efficacy of extracting information out of the complex neural data, and their employment for improving our understanding of neural networks, critically depends on the mathematical processing steps ranging from simple detection of action potentials in noisy traces - to fitting advanced mathematical models to distinct patterns of the neural signal potentially underlying intra-processing of information, e.g. interneuronal interactions. Here, we present a robust strategy for detecting signals in broadband and noisy time series such as spikes, sharp waves and multi-unit activity data that is solely based on the intrinsic statistical distribution of the recorded data. By using so-called "higher criticism" – a second-level significance testing procedure comparing the fraction of observed significances to an expected fraction under the global null – we are able to detect small signals in correlated noisy time-series without prior filtering, denoising or data regression. Results demonstrate the efficiency and reliability of the method and versatility over a wide range of experimental conditions and suggest the appropriateness of higher criticism to characterize neuronal dynamics without prior manipulation of the data.
\end{abstract}


\section{Introduction}

Advances in data acquisition technologies have led to the production of an enormous amount of complicated data sets in different scientific areas, which contain invaluable information. Thus, it is of great importance to develop and utilize statistical techniques that allow scientists to make new discoveries based on correct interpretations of data, rather than on false outcomes due to statistical insufficiency. Experimental tools and environment commonly induce noise in the measurements.  Preprocessing of data, with sequences of normalization, scaling, filtering and alike can often help, but always with a risk of losing information or inducing "featuring" that is nothing but analysis-effect. Thus, it is quintessential to use statistical methods that can detect the presence and exact location of signals in large noisy data sets. An effective way of performing this task is to use procedures of multiple hypothesis testing in large scale  \cite{Efr2004, Efr2011} to measure the likelihood that a detection has truly found an existing signal.

Higher criticism (HC) is an effective method that one can use to detect sparsely 
distributed weak signals in large statistical data sets \cite{DonJin2008, DonJin2015}. 
HC is a subtle statistical inference method since it is a {\it second level significance testing} for multiple comparisons, which is suitable for deciding whether a 
hypothesis for a data set can be rejected, cf. \cite{Hol2018}. Two 
important features of this method are the following. First, it has the ability to find the 
appropriate significance level for the rejection of a hypothesis, and, second, it allows one to localize in the data and detect the parts that were 
responsible for the rejection. Among the wide range of applications of HC, we 
mention its efficiency in detection of non-normality in astronomical data \cite{CayJinTre2005}, in 
genome-wide study of rheumatoid arthritis \cite{ParTriLeiHuBey2009}, and in thresholding for 
biomarker selection \cite{WehFra2012}.

Neurophysiological measurements for brain research inevitably use 
multiple electrodes simultaneously to record electrical activities of a 
number of neurons \cite{MitraMultiple}. 
This has raised the crucial need for devising spike sorting methods 
\cite{GreenValidation}, \cite{GreenAutomated}, which is a challenging problem and 
conspicuously lacks a consensus on best methods.  
In the present work, we show that HC can be used effectively to detect neurophysiological 
signals and large field deflections, and present a novel and robust method for peak sorting which is based on clustering and association of a threshold to HC values in each cluster.

\section{Methods}

\subsection{Animals}

Six adult male Wistar rats were used for the experiments. Animals were 
ordered from Charles River Laboratories (Sulzfeld, Germany) as specific 
pathogen free rats. Animals were pair housed in individually ventilated cages and on a 08:00 to 
20:00 dark to light cycle. Acclimatization period was at least 2 weeks. 
Room temperature was kept constant (23 $\pm$ 1 ?). Standard laboratory 
rat food and tap water were provided ad libitum. All animal experiments have been performed in 
accordance with the guidelines of the state of Baden-Wuerttemberg and 
have been approved by the local authorities.

\subsection{Electrophysiological measurements}

Animals were anesthetized with isoflurane (induction at $5\%$, 
maintenance at $1.5-2.5\%$), then positioned in a stereotaxic apparatus. 
Burr holes were drilled over the dorsolateral geniculate nucleus (dLGN: -4.2 AP, +3.6 ML).
Blue ($\lambda=453 \ nm$) and red ($\lambda=641 \ nm$) SemiLEDs Metal 
Vertical Photon Light Emitting Diode, clear lens LED light (luminous 
intensity: 1100mcd, chromaticity coordinates: 660, viewing angle: 100 
deg) was positioned centrally in front of the contralateral eye and the 
room lights were turned off. The light stimulus (highest level 
intensity) was delivered continuously at a rate of $1$ for 700 ms 
duration and 300 ms rest, and $20$ Hz for 35 ms duration and 15 ms rest. 
A 16-channel multi-wire custom designed recording electrode (NeuroNexus) 
was lowered into the dLGN (-4.0 DV). The electrode's final location was 
determined by electrophysiological verification that the recording 
contained cells characteristic of LGN cell firing evoked by stimulus 
light flashes. A silver wire inserted into the neck muscle was used as a 
reference for the electrodes. Electrodes were connected to a 
pre-amplifier (in-house constructed) via low noise cables. Analog 
signals were amplified by 5000 and filtered using an Alpha-Omega 
multi-channel processor (Alpha-Omega, Model: MPC Plus). Signals were 
then digitized at 25 kHz using a data acquisition device CED, Model: 
Power1401mkII). These signals were stored using Spike2 software (CED).

\subsection{Higher criticism and statistical inference}
\label{HCconcept}

HC provides a powerful statistical inference method 
to judge the existence of a signal in gathered data in favor of 
rejecting a hypothesis $H$.  The ordinary inference is based on 
using an $\alpha$-significance, where $\alpha$ is customarily taken 
to be $0.05$ (or a smaller number depending on the subject), 
and looking at the $p$-value $p=P(\tilde X | H)$, namely the 
probability of observing the data $\tilde X$ given the hypothesis $H$ is true.  
If $p$ is less than or equal to the significance level $\alpha$, we 
reject the hypothesis $H$, because if it were true, the observation 
of $\tilde X$ would be an unlikely event. 
Another way of saying this is that $\tilde X$ is {\it  significant at level} $\alpha$, 
cf. \cite{Hol2018}. 

Now imagine the data $\tilde X_1, \dots, \tilde X_m$ is gathered and one wishes 
to use a quantity to judge the validity of a hypothesis $H$. Tukey brought forth 
the HC quantity \cite{Tuk1994}
\[
HC_{m, \alpha} 
= 
\sqrt{m} \frac{ (\text{Fraction of significances at level $\alpha$}) 
- 
\alpha}{\sqrt{\alpha(1-\alpha)}}, 
\]
and suggested rejecting  the hypothesis if $HC_{m, \alpha}$ is greater 
than 2. Systematic studies of the subject have found more accurate 
thresholds for rejection, for example if the hypothesis is that 
the data has standard normal distribution, the decisive quantity is 
$\sqrt{2 \log \log m}$  (see Theorem 1.1 in \cite{DonJinHC2004}), 
which we shall  use in this paper.

\subsection{Higher criticism for non-normality test}
\label{Gaussianitytest}

We illustrate how one can use HC introduced in \S \ref{HCconcept} 
to test whether a statistical data set $\tilde X_1, \dots, \tilde X_m$ 
deviates from having a normal distribution, and to localize to the 
data points that are responsible for the non-normality. By definition, a random variable 
$\tilde X$  has the {\it normal distribution} with mean $\mu$ and standard deviation $\sigma$, 
denoted by $\tilde X \sim \mathcal{N}(\mu, \sigma^2)$,  
if its probability density function is given by 
$
f(x ; \mu, \sigma^2 ) =  \frac{1}{\sqrt{2\pi \sigma^2}} e^{-{(x-\mu)^2}/{(2 \sigma^2)}}.
$
The simplest normal distribution $\mathcal{N}(0,1)$ is called the {\it standard normal distribution}. 
Any random variable $\tilde X$ can be {\it standardized} by replacing it with $X = (\tilde X-\mu)/\sigma$, where 
$\mu = \mathbb{E}[\tilde X]$ and $\sigma = \mathbb{E}[(\tilde X - \mu)^2]^{1/2}$ are the mean and the standard deviation of $\tilde X$, respectively. The standardized version 
will be of mean 0 and standard deviation 1. It is a fact that the standardization of any normally distributed 
random variable has the standard normal distribution. 

Now we can explain how one can go through the following steps to use HC  
formulated in terms of $p$-values to 
locate parts of a data set $\tilde X_1, \dots, \tilde X_m$ that are responsible for its deviation from having a normal distribution:
 
\begin{enumerate}

\item  Standardize the data by setting 
$
X_i = (\tilde X_i - \mu)/\sigma,$ for $i =1, \dots, m,$ 
where $\mu$ is the mean and $\sigma$ is the standard deviation of 
 $\tilde X_1, \dots, \tilde X_m$. 

\item
Calculate the $p$-values by setting, for $i=1, \dots, m$, 
\[p_i= 
 P \left ( |\mathcal{N}(0,1)| > |X_i| \right)
= 
1- \frac{1}{\sqrt{2 \pi} } \int_{-|X_i|}^{|X_i|} e^{-x^2/2} \, dx.   
\]
This can be done efficiently by using the scipy.special package in 
python for values of the error function  
$
\textnormal{Erf} (x) = \frac{1}{\sqrt{\pi}} \int_{-x}^x e^{-t^2} \, dt.  
$
Namely, set 
$
p_i = 1 - \textnormal{Erf} \left (|X_i|/\sqrt{2} \right )$.  
To avoid division by 0 in the next step, if $p_i=1$ set it equal to 0.99999, 
and if $p_i=0$ set it equal to 0.00001.

\item 
Sort the $p_i$ in ascending order: 
$
p_{(1)} <  p_{(2)} <  \cdots < p_{(m)}.
$  

\item 
Let 
$
HC_{m, i} = \sqrt{m}\frac{i/m - p_{(i)}}{ \sqrt{p_{(i)}(1-p_{(i)})}},  
$
for $i=1, \dots, m$,
and find the maximum: 
$
HC_m =  \max_{1\leq i \leq m} HC_{m, i}.
$

\item 
If $HC_m$ is remarkably greater than 
$\sqrt{2 \log \log m}$, reject the hypothesis that the data has 
standard normal distribution,  find the indices $i$ between $1$ and $m$ for which $HC_{m, i}$ 
is greater  than a threshold, and localize at the corresponding data points to find the non-normal 
parts of the data. We will elaborate shortly on our method of finding  thresholds by clustering.

\end{enumerate}

\subsection{Signal detection in noisy time series} 
\label{peakdetection}
Given a time series $\tilde X_1, \dots, \tilde X_m$ whose signal is accompanied with Gaussian 
noise, in order to detect the signal, we use HC by going through the 
steps written  in the algorithm in \S \ref{Gaussianitytest}. The signal will be detected as the 
parts of the time series that cause it to have non-normal distribution along with the points that are 
in a small neighborhood of such points.

We recall from step 5 of the algorithm presented in \S \ref{Gaussianitytest} that the algorithm finds the non-normal 
parts of the time series as the points that correspond to the $HC_{m,j}$ that are greater than 
a threshold.  For the electrophysiological data in \S \ref{spikephysdataHCsec}, we choose 
different thresholds by $k$-means  clustering, using the  sklearn.cluster package in python,  
on the list of points $HC_{m,i}, i = 1,  \dots, m$.  The {\it silhouette score} measured 
with the sklearn.metrics package in python suggests the number of clusters, and we 
associate a threshold to each cluster by the following formula:  
\begin{equation}\label{formulaforthreshold}
mean(\textrm{cluster}) + \frac{1}{4} \left ( max(\textrm{cluster}) - min(\textrm{cluster}) \right ). 
\end{equation} 
These thresholds, when used according to the algorithm written in \S \ref{Gaussianitytest}, 
detect true signals in the electrophysiological data, and provide a method of sorting them 
based on their intrinsic statistical properties.

In order to illustrate the sensitive performance of the devised method with HC 
on non-normal data, we also calculate the {\it kurtosis} of the time series.  
The kurtosis of a random variable $\tilde X$ is by definition its fourth standardized moment, namely 
\begin{equation} \label{kurtosisformula}
Kurt(\tilde X)=\mathbb{E}\left [\left ( (\tilde X-\mu)/\sigma \right )^4 \right],
\end{equation}
where $\mu$ is the mean and $\sigma$ 
is the standard deviation of $\tilde X$.  In fact $Kurt(\tilde X)=\mathbb{E}[X^4]$, where $X$ is the standardization of 
$\tilde X$  (which was defined in \S \ref{Gaussianitytest}). Clearly $Kurt(\tilde X) = Kurt(X)$. 
It is known that the kurtosis of any normally distributed random variable is  
equal to 3. Therefore, a notable deviation from 3 in the kurtosis indicates that the time 
series has non-normal distribution. 
We use the scipy.stats package in python to calculate the kurtosis. Note that in this package, the 
command {\it kurtosis() } calculates the formula given by \eqref{kurtosisformula} and subtracts 
3 from the result.

\subsection{Simulations for detection criteria on number of data points for weak signals} 
\label{simulationidea}

\subsubsection{Detection of non-zero mean} 
\label{nonzeromeansimulation}
Given a statistical data set with the normal 
distribution $\tilde X \sim \mathcal{N}(\mu, 1)$ with $\mu \neq 0$, the objective is to find a criterion on $m$ 
such that HC will be able to detect that $\mu$ is non-zero. For this, we use the 
NumPy package in Python to generate data points $\tilde X_1, \dots, \tilde X_m$ whose distribution is 
$\mathcal{N}(\mu, 1)$, and calculate HC value $HC_m$ of the data by using the 
algorithm described in \S \ref{Gaussianitytest}.

Since there is randomness in the generation of the samples, in order to obtain a robust 
lower bound for $m$, we bootstrap the $HC_m$ values 100 times. We then analyze the bootstrapped 
$HC_m$ schematically and determine the number of necessary points for the detection of non-zero 
mean $\mu$ to be the value $m$ at which the curve of the bootstrapped $HC_m$ crosses over the 
curve given by $\sqrt{2 \log \log m}. $ The reason for the latter quantity is that it is a well-known fact 
(see for example Theorem 1.1 in \cite{DonJinHC2004}) that if the data has the standard normal distribution 
then $HC_m/\sqrt{2 \log \log m}$ approaches 1 in probability as $m \to \infty$, and deviations in the data 
from validity of this hypothesis manifest themselves in values for $HC_m$ that are greater than $\sqrt{2 \log \log m}$.

\subsubsection{Detection of sparse and weak signal}
\label{weaksparsesignaldetecsec}

Since we wish to use 
HC to detect weak signals in noisy data, it is a natural question to ask how many 
data points are necessary for finding out whether a small portion of the data has non-zero mean. 
We formulate this problem as follows. We assume that  a small portion of the data, whose sparsity 
is represented by some $\varepsilon >0$, 
has the normal distribution $\mathcal{N}(\mu,1)$ with $\mu \neq 0$ and the distribution of the rest 
of the data is $\mathcal{N}(0,1)$. It is natural to think that $\mu$ represents the intensity of 
the signal. Therefore, our objective is to find, depending on how small $\varepsilon$ and $\mu$ are, 
how many data points are necessary so that HC can detect the presence of a 
sparse and weak signal of this type.

Therefore we consider the random variable $\tilde X$ with 
the following distribution: 
\begin{equation} \label{epsilonmunormal}
\tilde X \sim \varepsilon \,\mathcal{N}(\mu, 1) + (1-\varepsilon) \, \mathcal{N}(0,1). 
\end{equation}
Since the signal 
in our data is independent of the noise, we can assume the two components 
in \eqref{epsilonmunormal} are independent, which based on basic properties of 
normal distributions yields: 
\[
\tilde X \sim \mathcal{N}(\mu, (1-\varepsilon)^2 + \varepsilon^2). 
\]
Using the NumPy package in Pyhton, we generate $m$ 
points $\tilde X_1,  \dots, \tilde X_m$ with this distribution 
and calculate its HC value $HC_m$ using the algorithm described in \S \ref{Gaussianitytest}. 
Due to the presence of randomness in generating the statistical samples for $\tilde X$, in order to 
obtain a robust result, we bootstrap the calculated $HC_m$ values 100 times. Then, for different 
choices of $\varepsilon$ and 
$\mu$, we study schematically  the bootstrapped $HC_m$ with respect to $m$. The decisive quantity for 
identifying the number $m$ of necessary points for the detection of the signal represented by 
$\varepsilon$ and $\mu$ 
with HC is the particular value for $m$ at which the bootstrapped $HC_m$ crosses over 
the curve given by $\sqrt{2 \log \log m}$.

\section{Results}

\subsection{Detection criteria for higher criticism} 
\begin{figure}[H]
\centering
\includegraphics[scale=0.3]{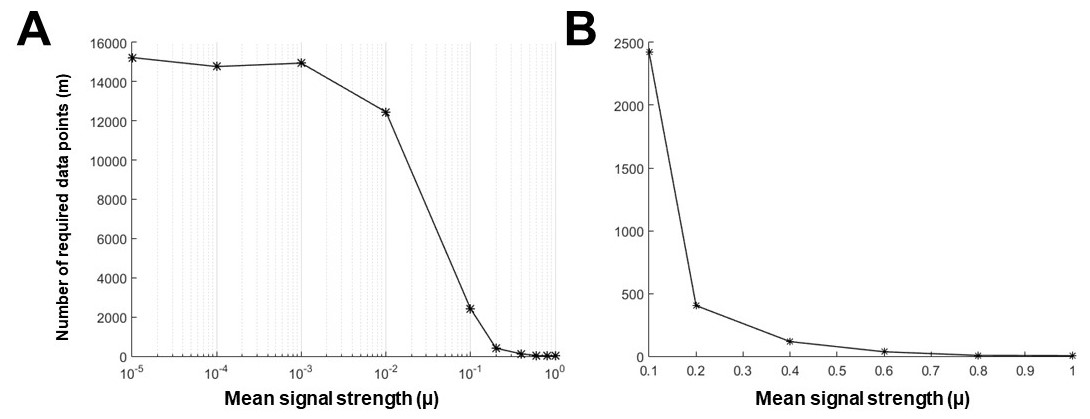}
\caption{\label{m_mu_fig2} 
Number of data points required for detection of the non-zero mean $\mu$  with higher criticism.}
\end{figure}

We report the results of our simulations explained in \S \ref{nonzeromeansimulation} 
in Fig. \ref{m_mu_fig2}:  the closer the mean $\mu$ to 0, the more data points 
are needed for the detection of the non-zero mean.

\begin{figure}[H]
\centering
\includegraphics[scale=0.34]{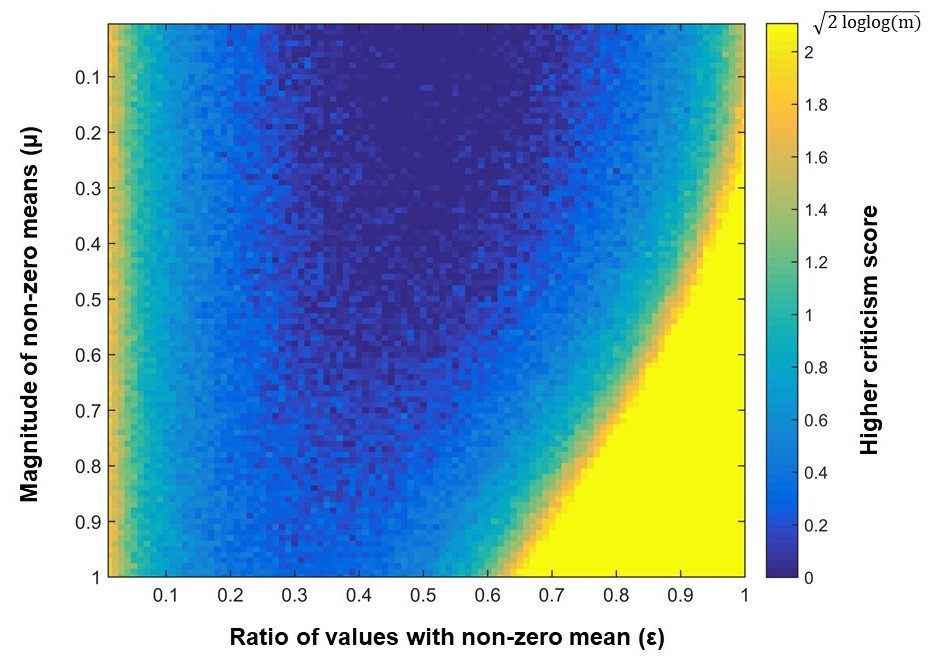}
\caption{\label{eps_mu_hc2} 
Number of data points required for detection of the weak (represented by $\mu$) 
and sparse (represented by $\varepsilon$) signal with higher criticism.}
\end{figure}

The results of our simulations explained in 
\S \ref{weaksparsesignaldetecsec} are presented in Fig. \ref{eps_mu_hc2}: 
more data points are needed for the detection of the signal when either 
its intensity decreases ($\mu$ closer to 0) or it is more sparse ($\varepsilon$ closer to 0).

\label{sparseandweaksimulation}

\subsection{Spike detection on electrophysiological data using higher criticism}
\label{spikephysdataHCsec}

\begin{figure}[H]
\centering
\includegraphics[scale=0.275]{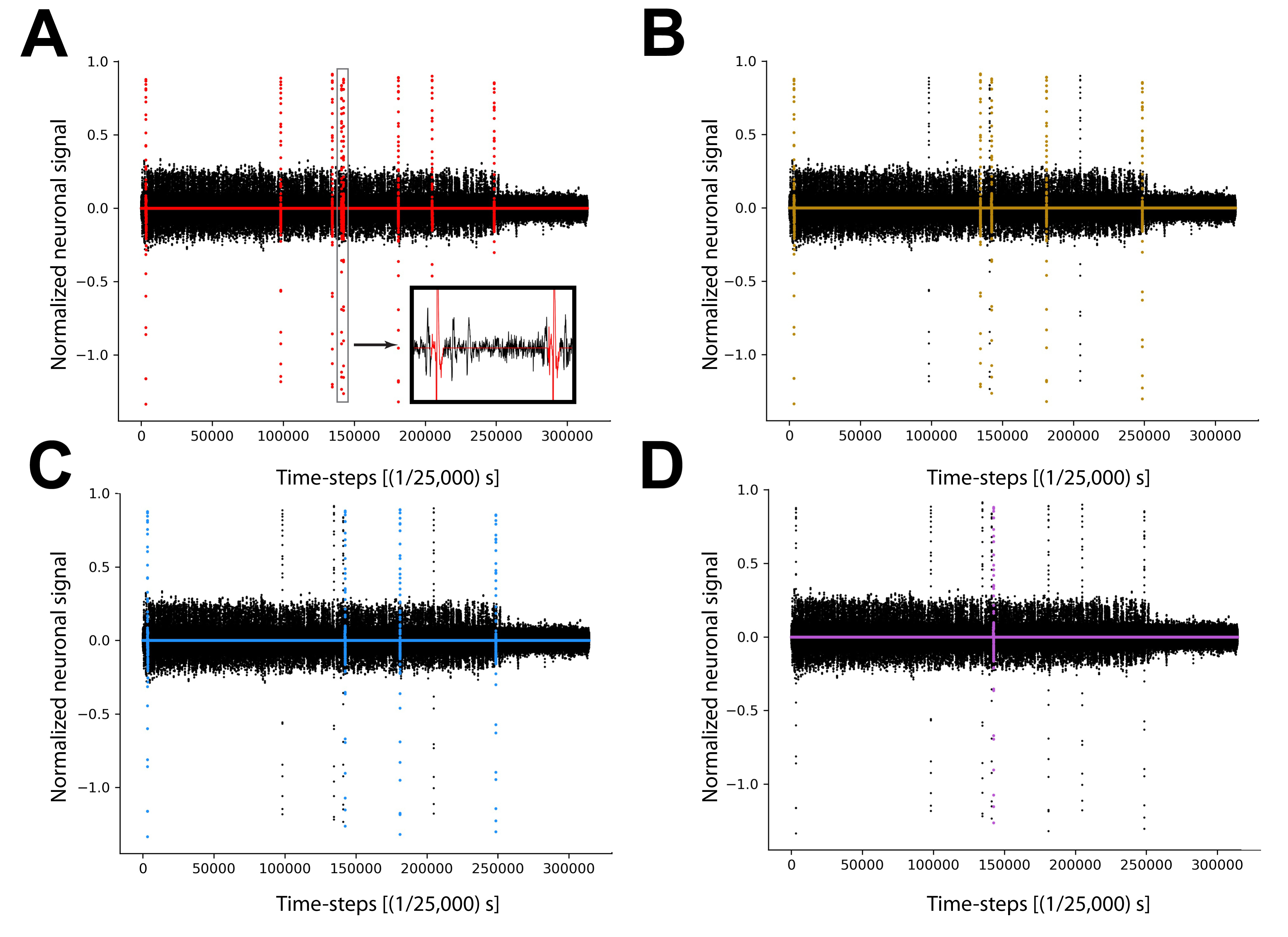}
\caption{\label{elctrophysspikesdetected} Neuronal signals (large field deflections) detected and sorted with higher criticism.}
\end{figure}

We explained in \S \ref{Gaussianitytest} and \S \ref{peakdetection} the details 
of our signal detection using HC. 
The kurtosis of the time series for the electrophysiological data shown in 
Fig. \ref{elctrophysspikesdetected} is equal to 
47.58, which indicates that it is far from having a normal distribution. Thus we standardize the 
time series and find the HC value $HC_m$ of the new series to be 
notably larger than $\sqrt{2 \log \log m}$, where $m$ is the length of the time series. The unusually large $HC_m$  is compatible with the fact that the kurtosis deviates considerably 
from 3, and they both assure that we should expect non-normality in the data.

In order to use the algorithm written in \S \ref{Gaussianitytest} to find the non-normal parts, we need to define a 
threshold for step 5 of the algorithm. We use different thresholds systematically 
as explained in \S \ref{peakdetection}: the 
silhouette score suggests that we divide the list of the $HC_{m,i}$ to 
4 clusters, and we associate a threshold to each cluster by \eqref{formulaforthreshold}. 
For each threshold, 
the points that correspond to the $i$ such that $HC_{m, i}$ 
is larger than than the chosen threshold allow us to localize to the source of 
non-normality in the data. When we plot these points along with the points that are 
within 50 time steps away from them, while flattening the remaining 
points to 0, we detect the signals (large field deflections) of electrophysiological time series, 
as shown in Fig. \ref{elctrophysspikesdetected}. That is, Fig. \ref{elctrophysspikesdetected}A 
shows the detection with the smallest threshold, where all and only the true signals are detected, 
with the ability of the method in recognizing two nearby large field deflections demonstrated. 
Fig. \ref{elctrophysspikesdetected}B, C, D respectively show the similar detection 
performed respectively with the 2nd, 3rd and 4th smallest thresholds. 
It is striking that as the threshold is changed with relatively big jumps, different 
but only large field deflections are detected, hence a sorting method for neuronal signals based on their statistical 
properties.

\section*{Acknowledgment}

FF acknowledges support from the Marie Curie/ SER Cymru II 
Cofund Research Fellowship 663830-SU-008. 
HRN received funding from the European Union's Horizon 
2020 research and innovation programme under grant agreement 
No 668863 (SyBil-AA) as well as the Bundesministerium für Bildung 
und Forschung (e:Med program: FKZ: 01ZX1503).

% when read from the biblio.bib
%\bibliographystyle{abbrv} 
%\bibliography{biblio}
%%%%%%%%%%%%%%

\end{document}